\documentclass[%
  reprint,
  superscriptaddress,
  amsmath,amssymb,D_1
  aps,
]{revtex4-1}

\usepackage{graphicx}
\usepackage{dcolumn}
\usepackage{bm}
\usepackage{mathrsfs}
\usepackage{amsmath}
\usepackage[usenames,dvipsnames]{color}
\usepackage{amssymb}
\usepackage{slashed}

\usepackage{color,soul}
\setul{0.5ex}{0.3ex}
\setulcolor{red}

\usepackage{subfigure}

\usepackage{hyperref}
\usepackage[mathlines]{lineno}

\begin{document}

\preprint{APS/123-QED}

\title{Small-scale structure from charged leptophilia}

\author{Ottavia Balducci}
\email{ottavia.balducci@physik.uni-muenchen.de}

\affiliation{Arnold Sommerfeld
  Center for Theoretical Physics, Theresienstra{\ss}e 37, 80333
  M\"unchen\\}
	  
\author{Stefan Hofmann}
\email{stefan.hofmann@physik.uni-muenchen.de}

\affiliation{Arnold Sommerfeld Center for Theoretical Physics,
  Theresienstra{\ss}e 37, 80333 M\"unchen\\}

\author{Alexis Kassiteridis}
\email{a.kassiteridis@physik.uni-muenchen.de}

\affiliation{Arnold Sommerfeld Center for Theoretical Physics,
  Theresienstra{\ss}e 37, 80333 M\"unchen\\}



\date{\today}

\begin{abstract}
We consider a charged leptophilic extension of the Standard Model of
particle physics as a minimal dark sector.  It accomodates a WIMP
paradigm at the TeV-scale that is sufficient to solve all small-scale
problems of $\Lambda$CDM and explain the excess of highly energetic
cosmic ray Standard Model electrons and positrons presented recently
by the DAMPE collaboration. The predictive power of this model allows
to test it in the near future.

\begin{description}
\item[PACS numbers] 04.20.Dw, 04.62.+v, 04.70.-s, 11.10.-z
\end{description}
\end{abstract}



\maketitle


\section{\label{sec:level1}Introduction}
In this work we are interested in the small-scale structures generated
by a gauge invariant charged leptophilic ultraviolet-complete
extension of the Standard Model (SM), which possibly explains the
excess of highly energetic cosmic ray SM-electrons and positrons (CRE)
presented recently by the DAMPE collaboration
\citep{Ambrosi:2017wek}. The CRE spectrum includes a tentative peak
excess around energies of 1.4 TeV.

This excess suggests that Dark Matter candidates should be able to
annihilate into SM charged leptons, especially into SM-muons and
SM-electrons \cite{Ge:2017tkd}. We study the thermal evolution of the
proposed dark sector (DS) and we develop a charged leptophilic theory
away from non-gauge invariant milli-charged couplings. We also study
the small-scale strucure of the proposed DM theory and we show that a
specific mass hierarchy setup is able to solve the missing satellite
problem together with the cuspy simulated profiles and the too big to
fail issue; for a thorough discussion about such theories see
\cite{Bringmann:2016ilk} and \cite{Tulin:2017ara}.  We point out that
since the DAMPE results \cite{Ambrosi:2017wek} became public, a whole
zoo of publications has appeared, e.g. \cite{Liu:2017rgs},
\cite{Chao:2017emq}, \cite{Ding:2017jdr}, motivating our work; however,
none of them was able to address the enduring small-scale problems of
$\Lambda$CDM studying the corresponding small-scale structure. This
research follows closely our previous neutrinophilic approach
\citep{Balducci:2017vwg} and the general study of the dark QED
paradigm \cite{Balducci:2018ryj} together with the work on small-scale
structure from neutron dark decays \cite{Karananas:2018goc}.\\

This paper is organized as follows. In Sec. \ref{sec:level1} we
explain the problematic of the construction of the desired partition
function and we introduce the proposed interactions in a gauge
invariant leptophilic extension of the SM.  Then, in
Sec. \ref{spectrum}, we compute various relevant observables in this
framework and present in Sec. \ref{parameters} a list of strong
constraints on the thermal evolution and the parameter set of the
theory. Then we test the minimal leptophilic theory in
Sec. \ref{evolution} to find out whether the small-scale issues are
cured successfully. Finally, in Sec. \ref{conclusion}, we conclude.

\section{\label{sec:level1}The charged leptophilia}
\subsection{Motivation of the framework}
In order to reproduce the peak excess in the CRE spectrum, we
introduce a DM candidate at the TeV-scale, which admits non-negligible
leptophilic interactions.  In other words, these interactions take
place between SM charged leptons and the DM particles and cannot be
mediated via the SM photon, due to strong constraints on milli-charged
theories.  Therefore, the stable DM particles can not carry
hypercharge and/or other SM-charges and vice versa: no SM-fields are
charged under any dark symmetries. In addition, we assume that the DM
candidate is a fermion following the stability of matter principle and
the need of astrophysically interesting properties
\cite{Tulin:2013teo}.

The only dimension-4 operator that realizes all the above
considerations is that of a Yukawa type, where the DM fermionic field
$e$ couples to a SM-charge lepton $l$. However, gauge invariance
demands the presence of an auxiliary dark field $\Delta^+$ carrying
exactly the opposite charges of $l$. For simplicity we assume that
$\Delta^+$ admits only a hypercharge of {$+1$} and therefore $l \in
(\textbf{1},\textbf{1},-1)$ is taken to be right-handed. This is the
same type of interactions arising from a supersymmetrical treatment of
the SM: a lepton-photino-slepton interaction; nevertheless, we do not
consider such extensions in our work, since we are only interested in
the consequences of this interaction on the small-scale structure.

\subsection{The partition function}
More precisely, the dark sector (DS) is based on a new global
symmetry, which for the sake of minimalism of the theory is taken to
be an Abelian U$(1)_r$.  As explained previously, the minimal
leptophilic theory is assumed to consist of a stable dark electron
field, which is charged under the global U$(1)_r$ with charge $r$ and
is a SM-singlet with mass parameter $M$. In other words, the $e$-field
can not be Majorana and has to be Dirac in accordance to the model
independent study in Ref. \cite{Athron:2017drj}. We call it dark
electron $e$, renaming the known right-handed charged lepton triplet
$l_R$ to SM-electron $e_R$, SM-muon $\mu_R$ and SM-tau $\tau_R$.  The
dark electron is accompanied by an auxiliary scalar doublet $\Delta
\equiv (\Delta^+, \sqrt{2}^{-1}\Delta^0)^{\rm T}$, where the first
field $\Delta^+$ is complex and admits a global charge $r$, a
$Y$-hypercharge {$+1$}, and a mass $M'$, while the second component
$\Delta^0$ is real and a total singlet with mass $M_0$. It is useful
to define $\mathbb{P}_\pm$, the up/down projections in the doublet
$\Delta$-space and the vector $\chi =(1,1)^{\rm T}$.This SM extension
includes modes electrically charged and therefore departs from the so
called ``nightmare scenario" of DM. Furthermore, in this work we are
agnostic about the origin of the masses of the initial theory.\\

The UV-complete partition function of the charged leptophilic theory
discussed above is given by
\begin{equation}\label{ds0}
Z_{r}= \int \mathcal{D}[\Psi_{\rm SM}, \Psi] \exp\{ \mathrm{i}
I[\Psi_{\rm SM}, \Psi]\}
\end{equation}
with $\Psi \in \{ e, \Delta \}$, $\Psi_{\rm{SM}}$ the SM-fields and $I$ the action functional using the following Lagrangian density
\begin{eqnarray}
\label{ds}
	\mathcal{L}_{\rm DS} &=& \bar{e}\mathcal{K}_{e} e +
        \Delta^{\dagger} \mathbb{K}_{\Delta}\Delta + \mathcal{L}_l+
        \mathcal{L}_1 \;.
\end{eqnarray}
Here, we used the useful kinetic kernel abbreviations $\mathcal{K}_{e}\equiv {\rm i}\slashed{\partial}-M$,
$\mathbb{K}_{\Delta}\equiv \mathbb{D}^{\dagger} \mathbb{D}-
\mathbb{M}^{\dagger} \mathbb{M}$ and the covariant derivative $\mathbb{D} \equiv
\mathbb{I}\partial -\mathbb{I}\, {\rm{i}}g'\, Y\, B$. Furthermore, we consider for simplicity mass eigenstates for the scalar field
$\mathbb{M}=\text{diag}(M',M_0)$; all mass parameters are positive and
$B$ is the SM-gauge boson associated with the hypercharge ($Y$).

The leptophilic virtue of the proposed theory is incarnated in the
minimal gauge invariant Yukawa potential discussed previously between SM right-handed
charged leptons, the dark electron and the auxiliary charged scalar field, namely
\begin{equation}
\mathcal{L}_l= -g_i \bar{e}\left(\chi^\dagger \mathbb{P}_+ \Delta \right) l^i_R + \text{h.c.}\;{},
\end{equation}
with $g_i \in \mathbb{R}$ and $i$ running over all the components of
the SM-charged lepton triplet.

Since the $\Delta^0$-field is a total singlet, further dimension-4
operators of purely dark interactions should be present in
$\mathcal{L}_1$; such terms, to the best of our knowledge, were not
included and therefore not studied in the leptophilic-related
literature so far. Here, we choose to present the most relevant ones
for the small-scale structure formation of the DM cosmology,
\begin{eqnarray}
\label{dsinte}
	\mathcal{L}_1 &\supset& - a_1\left(\Delta^{\dagger}
        \mathbb{P}_+ \Delta \right)\left(\Delta^{\dagger} \mathbb{P}_-
        \Delta \right) \nonumber{}\\ &&-\sqrt{2}\mu_+
        \left(\Delta^{\dagger} \mathbb{P}_+ \Delta \right)
        \left(\chi^{\dagger} \mathbb{P}_- \Delta \right)
        \nonumber{}\\ &&-\tfrac{\sqrt{2}}{3}\mu_0
        \left(\Delta^{\dagger} \mathbb{P}_- \Delta \right)
        \left(\chi^{\dagger} \mathbb{P}_- \Delta \right)
        \nonumber{}\\ &&+ \sqrt{2} g_0\, \bar{e} \left(\chi^{\dagger}
        \mathbb{P}_- \Delta\right) e\;{},
\end{eqnarray}
with $a_1, g_0, \mu_{+/0}/M' \in \mathbb{R}$.
At this point, it is proper to give some hints towards the necessity of each of the above interactions regarding the phenomenological signature of the proposed leptophilic framework. The first term is responsible for the relic density
depletion of the auxiliary charged scalars, the second one gives rise to the
scenario where the neutral scalar modes are unstable, the third one enables
IR-dominant, purely uncharged interactions, and the last one allows
attractive self-interactions between the DM modes: it is this very interaction that enables a self-interacting DM (SIDM) scenario leading to a phenomenologically viable theory with clear predictive power.

\section{The spectrum hierarchy and scatterings \label{spectrum}}
We continue by presenting the spectrum hierarchy of the theory, which
leads to extremely interesting phenomenological results due to a
spectrum degeneracy of the participating particles. Furthermore, we
compute important scatterings and decays of fields appearing in
the DS partition function (\ref{ds0}).

\subsection{The spectrum}
The dark electron $e$ and the charged scalar $\Delta^+$ admit masses at the
TeV-scale. Moreover, as we mentioned before, we entertain the possibility of the presence of
a very strong mass degeneracy, namely
\begin{equation}
M'-M=d>0\quad \text{with} \quad \delta \equiv\frac{d}{M}\ll 1\, ,
\end{equation}
but still $d> m_e$, where $m_e$ is the mass of the SM-electron; furthermore we take $d\ll m_\mu, m_\tau$. In this work, we consider the
benchmark mass $M = 1.4$ TeV to explain the peak-event of the DAMPE
CRE measurements. Therefore, $M'$ is fixed to 1.4 TeV up to
$\mathcal{O}(m_e)$ corrections.

On the other hand, the uncharged scalar $\Delta^0$ should admit a mass
well below the TeV-scale as explained in \cite{Tulin:2017ara}. Later on it will become clear that in order to obtain a viable
small-scale structure scenario $M_0 \sim \mathcal{O}(1)$ keV.

\subsection{Scatterings in the leptophilic theory}
We assume that $g_i\gg g_0$ in order to realize the electron/positron excess in the CRE spectrum.  This means that the dark electron modes annihilate
rapidly at $T<M$ into SM-charged leptons $l_i$ through the dominant
$s$-wave channel to dark radiation,
\begin{equation}
\langle v_{\rm rel} \sigma_{\rm ann} \rangle_{e\rightarrow i}
=\frac{\pi\alpha_i^2}{8 M^2} \sqrt{1-\delta_i'^2}\;{},
\end{equation}
where $\langle ... \rangle$ denotes the thermal average using relative
velocities, $\alpha_i \equiv g_i^2/4\pi$ and $\delta'_i \equiv
m_i/M$. Here, it becomes clear why a Majorana dark electron would disable
the $s$-wave channel and lead to $p$-wave suppressed annihilations
only, which is phenomenologically not acceptable \cite{Athron:2017drj}.

The annihilations of the charged scalars into dark electrons is of no
importance, due to the the heavy phase-space suppression, and their
annihilations into SM-electrons is $p$-wave suppressed. However, the
$\Delta^+$ particle number reduces mainly due to the annihilations
into uncharged scalars as long as $M'>M_0$ with a thermal averaged
cross-section of
\begin{equation}
\langle v_{\rm rel} \sigma_{\rm ann} \rangle_{\Delta^+\rightarrow
  \Delta^0} =\frac{\pi\alpha_1^2}{4 M'^2} \sqrt{1-\frac{M_0^2}{M'^2}}\;{},
\end{equation}
assuming $a_1 \gg (\mu/M')^2$ and $\alpha_1 \equiv a_1/4\pi$. Finally, their remaining negligible abundance is eliminated due to their rapid decays into SM-electrons and dark electrons.\\

We turn now our attention to the elastic scattering between the dark
electrons/positrons and the SM-electrons.  At temperatures $d, m_e \gg
T $ and at lowest order of perturbation theory, the momentum transfer
elastic cross-section, which is defined by $\sigma_{ T}\equiv \int
\mathrm{d}\Omega (1-\cos \theta) \frac{\mathrm{d}\sigma_{\rm
    el}}{\mathrm{d}\Omega}$, is approximated to
 \begin{eqnarray}
	   \sigma_T ^\mp \approx \frac{\pi \alpha^2_{{\rm
                 res}\,{}\mp}}{4M^2} \; .
\end{eqnarray}
In the above expression the possible resonant behavior is hidden in the effective coupling constant $\alpha_{\rm
  res} $, which is defined as
\begin{equation}
\alpha_{{\rm res}\,{}\mp} \equiv \alpha_e\, \delta' P_\mp(\delta,
\delta')^{-1}
\end{equation}
through the resonance polynomial in $\delta$ and $\delta'$
given by
\begin{equation}
P_\mp(\delta, \delta')=
-\left(\delta\pm\delta'(1+\tfrac{v^2}{2})\right)+
\tfrac{1}{2}\left(-\delta^2 + \delta'^2\right)\, ,
\end{equation}
with $\mp$ denoting the scattering of dark electron/positron modes on
SM-electrons and $v$ the SM-electron velocity in the rest frame of the
incoming dark electron/positron. Here, we omitted the lepton labels in
the $\delta, \delta'$ taking $i=e$, because we consider only
resonances near the SM-electron physical mass. One notices that as
$\delta$ approaches $\delta'$ the dark positron cross-section
$\sigma_T^+$ grows rapidly, which could lead to a kinetic freeze-in
and therefore to a possible late decoupling regime. However, for
temperatures of order $\mathcal{O}(1)$keV the SM-electron plasma
velocity is $v\sim 0.1$, which bounds the cross-section by the
SM-electron mass. The dark electron cross-section $\sigma_T^-$ could
be of similar strength but it is always bounded by the SM-electron
mass $m_e$. We note that the above realization is only valid as long
as
\begin{equation}
\vert P_\mp(\delta, \delta')\vert \gg \frac{4\pi \alpha}{9}
\left(\frac{T}{M}\right)^2\, ,
\end{equation}
which means that the thermal mass of the SM-electron in the SM-photon-plasma
\cite{Kalashnikov:1998av} is much smaller than the effective mass of
the corresponding $s/u$-channel mediator. Here, $\alpha$ is the SM
fine-structure constant.  During the above discussion we ignored the
optical term $ \Gamma/M' $ in the polynomial $P_\mp(\delta,\delta')$,
due to the total decay width $\Gamma$ of $\Delta^+$, since it is
negligible compared to the plasma temperature. In this work, we will also 
investigate, whether this resonant scattering suffices to provide solutions to at least some of the small-scale problems of the $\Lambda$CDM.\\

Even after the chemical freeze-out of the dark matter modes the scattering between them and the uncharged scalars
with energy $E$ becomes also of great importance, since it allows a
natural late decoupling scenario. More precisely, one obtains for the
the IR-dominant part of the momentum transfer elastic cross-section
\begin{equation}
\sigma_T \approx \pi \alpha_n^2 M^2\,
\frac{\log\left[\frac{4E^2-3M_0^2}{M_0^2}\right]-\frac{E^2-M_0^2}{E^2-\frac{3}{4}M_0^2}}{8(E^2-M_0^2)^2}\,
,
\end{equation}
where we defined $\alpha_n \equiv \frac{g_0\mu_0}{4\pi M}$. One
notices that as $T$ approaches the rest mass of the uncharged scalar,
the elastic scattering becomes extremely dominant.

The minimum leptophilic theory possesses all properties of a prototype
self-interacting dark matter theory (SIDM), which is realized through
the uncharged scalar mediator. The SIDM cross-sections per dark matter
mass, $\sigma_T$, are purely attractive and for given average
velocities are strongly velocity-dependent in the regime $ 2\alpha_0
M_0/(Mv_{\rm rel}^{\; 2}) \lesssim 10^3$, where $\alpha_0 \equiv
g^2_0/4\pi$. Such interactions appear to be a necessary ingredient in
order to resolve small-scale problems that are present during
structure formation in non-SIDM. The numerical solutions in various
regimes of the momentum transfer cross-sections can be found in
Refs.\cite{Tulin:2013teo}, \cite{Cyr-Racine:2015ihg}.\\

The charged scalar decays at tree-level
exclusively into dark electrons and positrons and SM-charged leptons (here only SM-electrons and positrons) due to the mass gap $d>m_e$ . This is described by the
decay width
\begin{equation}
\Gamma_{\Delta^+ \rightarrow e\, e_R } \approx \alpha_e \, \delta' \,
\sqrt{\delta^2+\delta'^2}\, M\, .
\end{equation}
This corresponds for example to a life time $\sim 10^{-12}$s for $M=1.4$ TeV, when
taking conservatively $d = m_e + 0.1\%$ with $\alpha_e =0.1$. \\

Concluding, we note that the uncharged scalar is in general unstable and decays, otherwise it would possibly over-close the
universe. Moreover, these scalars decay dominantly into photons
assuming $M_0<2 m_e$. The decay width is loop-suppressed and it is
given by
\begin{eqnarray}
	\label{Xsink}
	\Gamma_{\Delta^0 \rightarrow \gamma\gamma} &\approx &
        \mathcal{O}(1)\, \alpha^2 \left(\frac{M_0}{M'}\right)^4
        \left(\frac{\mu_+}{M_0}\right)\mu_+ \; .
\end{eqnarray}
For $T>M_0$, the inverse rate $\Gamma_{\Delta^0 \leftarrow
  \gamma\gamma}$ is a bosonic source, which is rescaled by a factor
$M_0/2T$ relative to the decay width.

\section{Parameters and constraints \label{parameters}}
A dark matter theory is always subject to various constraints in a
purely phenomenological approach. Therefore, in this section we
discuss such constraints on the parameter space.

\subsection{Particle physics bounds and detection}
The physical modes that interact directly with photons and SM-leptons
should admit a mass of at least $100$ GeV, considering the lowest
bound on neutralino and slepton masses \cite{Patrignani:2016xqp}. This
constraint does not restrict the proposed theory, since the dark electrons
and the charged scalars are assumed to be responsible for the 1.4 TeV CRE
excess measured by the DAMPE experiment.

We assume that all couplings lie in the perturbative domain. For the
dimensionful parameters $\mu_{+/0}$ this translates to $\mu_+ \lesssim M'$, 
which arises from the one-loop vertex correction of the mixed-scalar
cubic interaction, and $\mu_0 \lesssim M_0$. For example, the mass-squared correction of the uncharged scalar after
 the one-loop calculation due to this mixed interaction is given by
\begin{equation}
\Sigma_{\Delta^0}^+(q^2)/q^2 \approx - \frac{1}{192 \pi^2}
\left(\frac{\mu_+}{M'}\right)^2 \,
\end{equation}
for $q^2 \ll M'$. Therefore, assuming that the new physics regarding
the origin of the parameter $\mu_+$ is present not well above the mass
scale of the dark electrons and the charged scalars, we demand
conservatively $\frac{\mu_+}{M}\, ,\frac{\mu_0}{M_0} \ll 1$
as in \cite{Bringmann:2016ilk}.

The couplings appearing in the leptophilic Yukawa sector are subject
to further constraints. Firstly a WIMP condition should be present: as
we mentioned before, the dark electrons and positrons are thermally
produced WIMPs; therefore, the condition on the coupling constants
of the theory, $\alpha_i$, reads
\begin{equation}
\alpha_i > \left(\frac{M}{M_{\rm Pl}}\right)^{1/2}\, .
\end{equation}
It is important to note that the proposed new interactions between SM-charged leptons
should eventually contribute to the anomalous magnetic moment of the SM-lepton. We
calculated the corresponding factor due to the one-loop vertex
correction at zero momentum transfer as
\begin{equation}
\lim_{q^2\rightarrow 0}\delta F_2^i= -\frac{\alpha_i}{48\pi}
\delta_i'^2\, .
\end{equation}
The above contribution should be less than the measured values for the
leptonic Land\'e-factors, $\frac{g-2}{2}$, found in
\cite{Patrignani:2016xqp}. Furthermore, one notices that any perturbative couplings
are admissible for all the SM-leptons. If, on the other hand, parity
were conserved and the Yukawa portal couples also to left-handed
SM-leptons in a similar manner, then the correction would be $\frac{\alpha_i}{12\pi}
\delta_i'$
and the coupling constants are severely suppressed. In other words,
the gauge invariant chiral portal protects the interaction strength
and allows the existence of thermally produced WIMPs at the TeV-scale.

Moreover, the presence of the charged scalar induces an effective
magnetic moment for the dark electron of order $\mu_e \approx
10^{-5}\mu_p$, leading to a spin-dependent cross-section
\cite{Dvorkin:2013cea}
\begin{equation}
\sigma_{\rm{proton}} \approx 3\alpha \mu_e^2 \sim \alpha^2\,
\sigma_{{\rm{ann}}\, e\rightarrow e_R} \sim
\mathcal{O}(10^{-40})\rm{cm}^2\, ,
\end{equation}
where for simplicity we assumed that $\alpha_i = \alpha_e \,
\delta_{e\, i}$. A similar problematic applys to the spin-independent
cross-sections. Such cross-sections are in line with the LUX and XENON
indirect detection measurements of WIMP-proton elastic scatterings
\cite{Hamze:2014wca}, \cite{Akerib:2016vxi}, \cite{Chao:2016lqd},
\cite{Aprile:2017iyp}.\\

Ending the discussion about the present bounds on parameters which
arise from particle physics, we note that for $M, M' \approx 1.4$ TeV
no constraints on the coupling constant $\alpha_e$ can be inferred
from LEP measurements \cite{Freitas:2014jla}; however, the TeV-scale
should be directly testable in the new ILC experiment
\cite{Freitas:2014jla}.

\subsection{Bounds on light degrees of freedom}
In this work, the benchmark point given by $M = 1.4$ TeV and $M_0 \sim
\mathcal{O}(1)$ keV is of great interest; these light modes of the
uncharged scalars decouple from the SM-plasma when the heavy particles
become non-relativistic at $T\sim M$. However, if these modes
freeze-in due to IR-dominant processes before the SM-neutrinos
decouple, this could modify the interactions responsible for the big
bang nucleosynthesis (BBN). Examining the parameter set of the
leptophilic theory, we find that for perturbative couplings this
interference is not expected to be realized. More precisely, we study
the deviation of the effective SM-neutrino degrees of freedom,
parametrizing the relativistic energy budget of the universe following
\cite{Balducci:2017vwg}.

For example, for $M_0 \sim \mathcal{O}(1)$ keV
the BBN processes are not affected and only around $x\sim 1$ with $x :=
M_0/T$ the SM-photon-plasma cools down and during the BBN period we
obtain $\Delta N_{\rm eff}\vert_{\rm BBN} \approx 0.03$. Nevertheless,
after the decay of the uncharged scalars into photons, the
photon-plasma experiences a reheating and the final deviation of the
effective SM-neutrino degrees of freedom can be estimated as $N_{\rm
  eff}\approx 2.97$ using equilibrium physics as first order
approximation. Concluding, we note that both results are not only
perfectly compatible with BBN \cite{Cyburt:2015mya} and CMB
\cite{Ade:2015xua} 1$\sigma$ measurements but they may lead also to a
solution for the recent tension regarding the decrease of $\Delta
N_{\rm eff}$ from BBN to CMB-based measurements, $\Delta N_{\rm
  eff}\vert_{\rm CMB}- \Delta N_{\rm eff}\vert_{\rm BBN}<0$. Finally,
if $M_0$ is less than the recombination temperature, then $\Delta
N_{\rm eff}\vert_{\rm CMB}=0.03$, which is still CMB
1$\sigma$-compatible.

\section{Thermal evolution towards a solution to the DM problem \label{evolution}}
Until this point, we defined the complete framework of the proposed theory and
we set the most recent bounds on the corresponding parameter space.
Now we are ready to compute  the DM relic density, which consists of electrons and
positrons and to study their kinetic decoupling behavior. All the
temperatures are given in the SM-photon-plasma frame.

\subsection{The present DM density}
In this section the following assumptions has to be made for simplicity:
the DM relic abundance consists
mostly of the dark electrons and dark positrons. We thus neglect
the relic density of the charged scalar {$\Delta^{+}$}, i.e.  no mixed
dark matter is present. This assumption is naturally accommodated in our theory for the given spectrum hierarchy, since
the {$\Delta^{+}$} scalars decay before the dark electrons and
positrons acquire their relic abundance.

In order to compute the relic density, we followed the same procedure as in
\cite{Balducci:2017vwg}, i.e. we solved the Boltzmann equation numerically requiring that the dark modes depart significantly from their
equilibrium distribution at the time of the chemical freeze-out.  The
DM relic density is found to be
\begin{eqnarray}
	\label{om}
	\Omega_{e^\mp} h^2 &\approx& \frac{1}{2}\times{}0.12\;
        \left(\frac{\alpha_l}{0.1}\right)^{-2}\; \left(\frac{M}{{1.4\,
            \rm TeV}} \right)^2\;{}, \;
\end{eqnarray}
with the leptophilic coupling constant {$\alpha_l\equiv \sqrt{\sum_i
    \alpha_i^2}$}. The 1/2 factor in front is due to the assumption of
a symmetric $e^-,e^+$ ensemble. One finds that the total annihilation
cross-section is of order $\sim 10^{-26}\rm{cm}^3 \rm{s}^{-1}$, which
can explain nicely not only the CRE peak of the DAMPE spectrum but
also the sizable hidden excess in the DAMPE measurements between 0.6
TeV and 1.1 TeV \cite{Ge:2017tkd} if the proposed flavor structure is
adopted. The constraints and the relic abundance (\ref{om}) fix the
leptophilic coupling constant $\alpha_l$ completely but not the
generation-oriented ones, i.e. $\alpha_i$.

\subsection{Multiple kinetic decoupling regimes}
There are mainly two different mechanisms in the elastic scattering
history of the dark electrons and dark positrons in this minimal leptophilic framework: firstly, the dark positrons are able to scatter
resonantly with the SM-electrons, and, secondly, all DM
particles could interact efficiently with the uncharged scalars through an
IR-dominant $t$-channel process.

We start by studying the first case, assuming that the IR-dominant scalar channel is absent: the elastic scattering of
the dark positrons with the SM-electrons. As long as the SM-electrons become non-relativistic the dark positrons depart from their kinetic
equilibrium with the SM-electrons and reenter this regime only near the resonance of the
corresponding elastic cross-section {$\sigma_{T}^{+}$}. More precisely, the situation appearing in the thermal
evolution of the proposed theory is somewhat more intricate: not all
dark thermal relics are in LTE at later times with the SM-electrons,
only the dark positrons admit this virtue and only for a limited time
window; furthermore, the SM-electrons are already thermal relics
themselves after their annihilation to SM-photons took place around
$T\sim m_e$. Therefore, the SM-electron density is characterized by the present photon to baryon ratio $\eta$ assuming an electrically neutral universe after the BBN epoch as in
\cite{Dvorkin:2013cea}, \cite{Gondolo:2016mrz},  and, technically, one can estimate the momentum transfer scattering rate \cite{Gondolo:2016mrz},
using the following
expression
\begin{equation} \label{kinetic decoupling}
\Gamma_{+\, {\rm el}}(T)\approx \frac{5\, \eta}{\pi} \frac{ \delta'}{ 1+\delta'}\,  \frac{\sigma_T^+(\delta,\delta')}{\sqrt{m_e}} \, T^{7/2} \, .
\end{equation} 

Equating the above rate with the Hubble one, we obtain the approximate
kinetic decoupling temperature $T^+_{\rm kd}$. As a benchmark point we
choose thermal relics with $M=1.4$ TeV and $d = m_e + 0.1\%$, or,
equivalently, $\left(\min\left[\vert P_+(\delta,\delta')\vert
  \right]\vert_{T=420\, \rm{eV}} \right)^{1/2} M \approx 720$ keV $\gg
T_{\rm BBN}$. This parameter set yields a temperature for the kinetic
decoupling $T_{\rm{kd}}^{+}\approx{} 420$ eV. For this choice of
parameters the dark positrons and SM-electrons are in kinetic equilibrium
for around two hours. Furthermore, we notice that $T^+_{\rm kd}$ is
extremely sensible to small changes of the mass gap $d$ between $M'$
and $M$, but it changes slowly as $M$ runs in the TeV-regime. This
indicates a possibly unstable late kinetic decoupling: for
temperatures slightly higher than $T^+_{\rm kd}$ the dark positrons may not
be coupled to the SM-electrons at all, due to the tiny value of
$\eta$.  Indeed, one finds out that these two hours are not sufficient  for the DM modes to enter a kinetic equilibrium with the SM-electrons ruining the hopes for a viable late kinetic decoupling regime of leptophilic DM theory even through a resonant scattering channel; quantitatively only one out of ten DM particles experiences a collision during that time interval.\\

However, this leptophilic theory provides an alternative way towards
$\mathcal{O}(1)$keV values of kinetic decoupling termperatures: the
elastic scattering of the DM particles with the uncharged scalars,
which turns out to be stable enough. The dark electrons and dark
positrons are in LTE with the $\Delta^0$-particles at high
temperatures; when the temperature falls below the rest mass of the
uncharged scalar field, it annihilates and decays into photons. The
main consequence is the rapid reduction of the $\Delta^0$-density,
which makes the remaining elastic scatterings between the DM particles
and the uncharged scalars exponentially suppressed. The rate for the
elastic scattering with {$\Delta^{0}$} can be found numerically using
the rate of the averaged momentum cross-section as in
\cite{Bringmann:2016ilk} and is
\begin{equation}
\Gamma_{0\, {\rm el}}(T)\approx \frac{1}{3\pi^2 M}\int_{M_0}^\infty
\mathrm{d}E\, f_{\Delta^0}(E) \frac{\partial}{\partial
  {E}}\left((E^2-M_0^2)^2 \sigma_{ T}\right)\,{}.
\end{equation}
Here, $ f_{\Delta^0}(E)$ is the distribution function per d.o.f. of
the uncharged scalar field at temperature $T$. In order to approximate
the time when the these elastic scatterings cease to be efficient, we
equate as before the elastic scattering rate with the Hubble
one. Therefore, the kinetic decoupling temperature $T_{\rm kd}^{0}$ is
to be found.  It is interesting to note that $T^0_{\rm kd}$ is not
very sensible to small changes of $M_0 \sim \mathcal{O}(1)$ keV.  For
the benchmark point $M_0 = 7.1$ keV, $\alpha_0 = 1.3 \times 10^{-4}$,
and $\mu_0 = \frac{1}{22}M_0$ we obtain $ T^0_{\rm kd} \approx 420$
eV.

\begin{figure}
\includegraphics[scale=1]{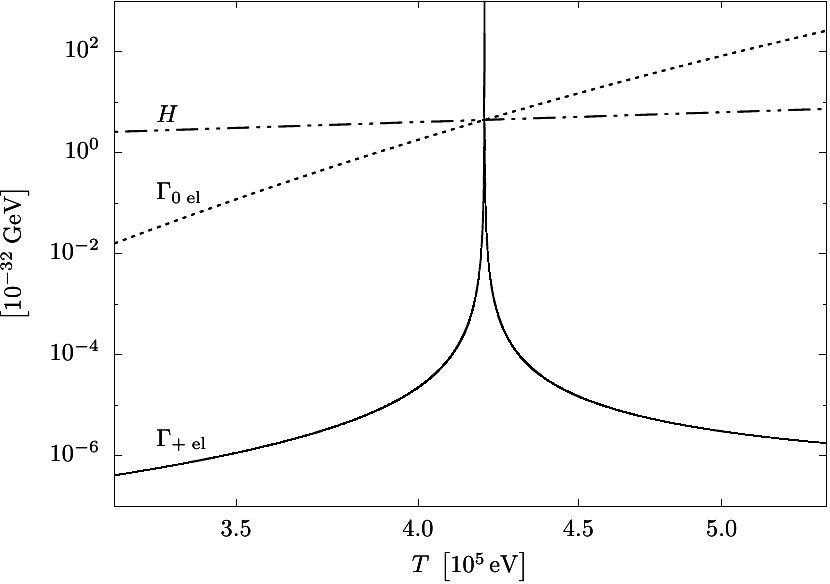}
\caption{The late kinetic decoupling regime at sub-keV
  temperatures. We plot the relevant cosmological rates (in units of
  $10^{-32}$ GeV) over the photon temperature (in units of {$10^{5}$}
  eV): the Hubble rate, $\Gamma_{+\, {\rm el}}$, and $\Gamma_{0\,
    {\rm el}}$.}
\label{graph1}
\end{figure}

Finally, we present the above discussed rates and results for
the given parameter set in Figure \ref{graph1}.
 We remark that smaller values of $M_0$ give rise to lower
decoupling temperatures, thus depleting the small-scale structure of
dark matter and leading to unacceptable results.\\

\subsection{The small-scale structure}
A late kinetic decoupling has positive effects for structure formation
at small scales: it yields easily the correct protohalo masses and
helps alleviating the cusp vs. core and too big to fail
problems as well, as argued in \cite{Schewtschenko:2015rno} and
\cite{Vogelsberger:2015gpr}. 
Moreover, for optimal results  the kinetic decoupling should happen after BBN following the discussions in \cite{Balducci:2017vwg}, \cite{Balducci:2018ryj}, \cite{Karananas:2018goc}, but long before the recombination period, in order to satisfy the strong Lyman-{$\alpha$}
constraints \cite{Baur:2015jsy}, \cite{Irsic:2017ixq}. 
In other words, the usual solution to the missing satellite problem of
$\Lambda$CDM-cosmology can be found after suppressing the linear power spectrum at
scales as large as that of dwarf galaxies \cite{Vogelsberger:2015gpr},
\cite{Boehm:2000gq}, \cite{Aarssen:2012fx}, i.e. for damping masses
corresponding to the desired temperatures of the late kinetic decoupling.

Furthermore, the combination of the SIDM effects, which lower the
inner densities of the galaxies energizing the particle modes,
together with the suppression of the linear power spectrum of the
leptophilic theory leads to a possible cure of the two remaining
small-scale issues of $\Lambda$CDM-cosmology in the non-linear regime:
the cusp vs. core and the too big to fail problems.  Until now, there
was no leptophilic theory able to explain the CRE excess in the DAMPE
spectrum and at the same time yield the correct small-scale structure.

\section{Results, discussion and conclusion \label{conclusion}}

The constrained parameter set of the proposed theory includes
naturally values of SIDM elastic cross-sections of order $ \sim 0.1-10
\, {\rm cm}^2 {\rm g}^{-1}$ \cite{Bringmann:2016din} as long as $M_0
\sim $ keV and $\alpha_0 \sim \mathcal{O}(10^{-4})$. We choose a benchmark point for
thermally produced WIMPs with $M=1.4$ TeV, $M_0=22 \mu_0= 7.1$ keV,
and a singlet coupling $\alpha_0 = 1.3 \times 10^{-4}$; therefore, a late kinetic
decoupling scenario exists independently from possible resonances, namely $T_{\rm kd} \approx$ 0.42 keV, and
at the same time the thermally averaged velocity dependent SIDM
cross-sections are similar to those of the tuned ETHOS-4 model
\cite{Vogelsberger:2015gpr}. This model is known in the DM community to be compatible with all present constraints
and alleviates the missing satellite, the too big to fail and
also the cusp vs. core problems \cite{Lovell:2017eec}.

In Ref. \cite{Bulbul:2014sua}  a mysterious photon-ray
measurement at 3.55 keV is presented, which indicates the possible existence of lighter
particles (well below the TeV-scale) in the DM spectrum. This line is naturally
 accommodated  in our model, since the
scale for the decaying $\Delta^0$ seems to be the keV-scale in order to alleviate the small-scale abundance problem. 
In other words, the minimal leptophilic theory does
not only explain the CRE excess at the TeV-scale in the DAMPE spectrum
\cite{Ambrosi:2017wek}, but it also favors at least one much lighter
particle to cure all the small-scale problems simultaneously, delivering damping masses and SIDM
cross-sections of the desired order and predicting
the cold DM present abundance.\\

In this paper we showed for the first time that charged leptophilic
dark matter candidates could solve all enduring small-scale problems
of the $\Lambda$CDM-cosmology and at the same time stay in line with
all recent astro-physical experiments. The proposed theory includes a
gauge invariant Yukawa portal, which couples the dark electron field to the
SM right-handed electron one. Two auxiliary fields are needed: a heavy
charged scalar and a much lighter uncharged scalar. However, even if a strong mass degeneracy between the dark electron and the charged scalar
is present, the delivered small-scale structure is unacceptable if no uncharged scalars are included in the theory. Therefore, the existence of the uncharged scalars is necessary to realize the SIDM nature of the
theory together with a stable conventional late elastic scattering
regime.  This theory enables a parameter space of the tuned ETHOS-4 model \cite{Vogelsberger:2015gpr},
\cite{Lovell:2017eec} and the properties of late decoupling and magnitudes of SIDM
cross-sections are able to potentially solve the enduring small-scale
problems of $\Lambda$CDM: the satellite abundances, the cuspy profiles
and the massive subhalos.  In addition, this minimal charged
leptophilic UV-complete extension of the Standard Model (SM) possibly explains the excess of highly energetic cosmic
ray SM-electrons and positrons (CRE) presented recently by the DAMPE
collaboration and its predictive power allows further testing
in the near future.

\begin{acknowledgments}
We appreciate financial support of our work by the DFG cluster of
excellence 'Origin and Structure of the Universe' and the Humboldt
Foundation. O.B. is grateful for financial support from the
'FAZIT-STIFTUNG'.

\end{acknowledgments}


 

\end{document}